\documentclass[reprint, superscriptaddress,amsmath,amssymb,prl,floatfix,showpacs]{revtex4-1}

\usepackage[FIGTOPCAP,tight]{subfigure}
\usepackage{graphicx}
\usepackage{dcolumn}
\usepackage{bm}

\usepackage{xcolor}

\begin{document}

\preprint{APS/123-QED}

\title{Measurement of sub-pulse-width temporal delays via spectral interference induced by weak value amplification}

\author{Luis Jos\'{e} Salazar-Serrano}
\affiliation{ICFO-Institut de Ciencies Fotoniques, Mediterranean
Technology Park, 08860 Castelldefels (Barcelona), Spain}
\affiliation{Physics Department, Universidad de los Andes, AA
4976, Bogota DC, Colombia}

\author{Davide Janner}
\affiliation{ICFO-Institut de Ciencies Fotoniques, Mediterranean
Technology Park, 08860 Castelldefels (Barcelona), Spain}

\author{Nicolas Brunner}
\affiliation{Departement de Physique Theorique, Universite de
Gen\`{e}ve, 1211 Gen\`{e}ve, Switzerland} \affiliation{H.H. Wills
Physics Laboratory, University of Bristol, BS8 1TL, United
Kingdom}

\author{Valerio Pruneri}
\affiliation{ICFO-Institut de Ciencies Fotoniques, Mediterranean
Technology Park, 08860 Castelldefels (Barcelona), Spain}
\affiliation{ICREA-Instituci\'o Catalana de Recerca i Estudis
Avan\c cats, 08010 Barcelona, Spain}

\author{Juan P. Torres}
\affiliation{ICFO-Institut de Ciencies Fotoniques, Mediterranean
Technology Park, 08860 Castelldefels (Barcelona), Spain}
\affiliation{Universitat Polit\`{e}cnica de Catalunya, Barcelona
Tech, Dept. of Signal Theory \& Communications, 08034 Barcelona,
Spain}


\begin{abstract}
We demonstrate experimentally a scheme to measure small temporal
delays, much smaller than the pulse width, between optical pulses.
Specifically, we observe an interference effect, based on the
concepts of quantum weak measurements and weak value
amplification, through which a sub-pulse-width temporal delay
between two femtosecond pulses induces a measurable shift of the
central frequency of the pulse. The amount of frequency shift, and
the accompanying losses of the measurement, can be tailored by
post-selecting different states of polarization. Our scheme
requires only spectrum measurements and linear optics elements,
hence greatly facilitating its implementation. Thus it appears as
a promising technique for measuring small and rapidly varying
temporal delays.
\end{abstract}
\maketitle

The measurement of temporal delays between optical pulses is
essential in metrology, for instance for accurate distance
measurements and for timing synchronization
\cite{kim2008,lee2010}, where the capability of discriminating
between small temporal delays with a {\em reference} pulse is
needed. Diverse optical schemes for measuring subpicosecond
temporal delays have been demonstrated. This is the case, for
instance, of schemes based on the use of ultrafast nonlinear
processes such as second harmonic generation
\cite{trebino2005,avi2005} or two-photon absorption
\cite{boitier2005}.

In another context, the well-known Hong-Ou-Mandel effect makes use
of quantum interference to measure subpicosecond temporal delays
between photons \cite{hom1987}, which was used by Steinberg et al.
\cite{steinberg1993} for measuring very small single-photon
tunneling times. Since this technique is based on measuring
two-photon coincidences, it generally restrict the number of
photons of the signal. However, quantum-inspired interferometers
\cite{kaltenbaek2010} might broaden the applicability of quantum
concepts to other scenarios.

When two similar optical pulses with temporal width $\tau $, and
time delay $T \gg \tau$ between them, recombine, a modulation of
the spectral density appears
\cite{alford_gold1958,givens1961,mandel1962}, which allows
measuring the time difference $T$. This is true even if the
optical path difference is larger than the coherence length of the
pulses \cite{zou_mandel1992}. However, for small values of T ($T
\ll \tau$), inspection of the spectral density reveals no
interference effects, even though interference manifest now in the
temporal domain as a periodic change of the output intensity as
function of the delay.

Here we demonstrate experimentally a scheme to measure small
temporal delays $T$ between optical pulses, much smaller than the
pulse width $\tau$, based on an interference effect in the
frequency domain which produces a measurable shift of the central
frequency of the pulse \cite{brunner2010}. The scheme makes use of
linear optics elements only and works in both the high and low
signal regimes. It allows the measurement of temporal delays
between optical pulses up to the attosecond timescale
\cite{strubi2013}. This phenomenon, which is inspired by the
concepts of quantum weak measurements and weak value amplification
\cite{aharonov1988,duck1989,aharonov1990,boyd2013}, produces interference
effects in the regime $T \ll \tau$, which allows to deduce the
value of $T$.

Although the concept of weak measurements originates from research
on quantum theory, the phenomenon of weak value amplification can
be readily understood in terms of constructive and destructive
interference between waves \cite{duck1989,howell2010,torres2012}.
In a weak measurement scenario, a system is weakly coupled to a
pointer (the measuring device). While the weakness of the coupling
can be seen as a disadvantage at first sight, Aharonov and
colleagues \cite{aharonov1988} showed that when appropriate
initial and final states of the system are selected (i.e. pre- and
post-selection), the pointer is shifted by an unexpectedly large
amount. It was soon suggested that these ideas may find
application in metrology \cite{aharonov1990,boyd2013}. This phenomenon,
termed {\em weak value amplification}, have been demonstrated
experimentally \cite{ritchie1991,brunner2004,pryde2005}, and have
been used for measurements of very small transverse displacements
of optical beams \cite{hosten2008,ben_dixon2009}, as well as for
frequency \cite{starling2010} and velocity measurements
\cite{viza2013}. Techniques for measuring small phase shifts have
also been proposed \cite{brunner2010,li_guo2011,strubi2013}.

\begin{figure}\label{fig:SuperModulatorTimeShift}
       \includegraphics[width=0.45\textwidth]{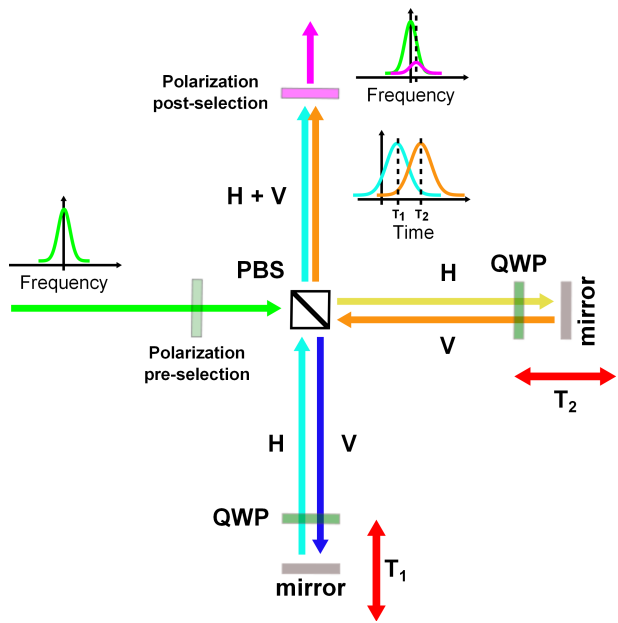}
\label{figure1} \caption{Schematic of the experimental setup.
State pre-selection: The polarization of the input optical pulse
is selected by using $\lambda/2$ and  $\lambda/4$ wave-plates (not
shown). Weak coupling: A Michelson-Morley interferometer, composed
of a Polarizing Beam Splitter (PBS), two $\lambda/4$ wave-plates
and two mirrors, divides the input pulse into two pulses, with
equal power and with orthogonal polarizations, that travel through
different paths of the interferometer. A movable mirror mounted on
a translation stage in one of the paths allows changing the
temporal delay. State post-selection: The two pulses recombine in
the PBS, and they are projected into a particular state of
polarization with a electrically-controlled liquid crystal
variable retarder (LCVR) and a polarizer. The output beam is
finally focused in a single mode fiber (SM) and its spectrum is
measured with an Optical Spectrum Analyzer (OSA).}
\end{figure}

Here, the weak coupling is realized by means of a
polarization-dependent temporal delay implemented in a Michelson
interferometer configuration (see Fig. 1). Brunner and Simon
\cite{brunner2010} showed that the introduction of a small
temporal delay between the two components (horizontal and
vertical) of a circularly-polarized pulse, can yield a large
central frequency shift after recombining the pulses and
projecting them into a polarization state nearly orthogonal to the
input state. However, the near orthogonality of the input and
output polarization states introduces heavy losses. Nevertheless,
the weak value amplification can also be used when the input and
output polarization states have a relatively large overlap, hence
away from the usual weak value amplification regime
\cite{torres2012}, allowing for the observation of significant
frequency shifts without heavy losses, as we will demonstrate
here.

Recently, it has been demonstrated that high precision phase
estimation based on weak measurements can be achieved even using
commercial light-emitting diodes \cite{xu_guo2013}. Indeed, Li et
al. \cite{li_guo2011} showed that the scheme proposed by Brunner
and Simon also works with large-bandwidth incoherent light. On the
one hand, the use of white light allows to obtain in a
straightforward manner a light source with an enormous bandwidth,
which allows to measure very small phase differences. On the other
hand, many applications make use of high-repetition femtosecond
sources that allows to perform multiple measurements in
millisecond or microsecond time intervals \cite{lee2010}, allowing
the measurement of time-varying phase differences in this time
scale. This is the scenario that we consider here.

We make use of a femtosecond fiber laser (Calmar Laser -
Mendocino) centered at $1549\,\mathrm{nm}$ (temporal width:
$320\,\mathrm{fs}$; average power: $3\,\mathrm{mW}$; repetition
rate: $20\,\mathrm{MHz}$). The spectral density measured shows
characteristic high-frequency small wrinkles due to cavity effects
in the laser system. The spectral density is $S_{in}(\nu)=1/2\,\,
\epsilon_0 c |E_{in}(\nu)|^2$, where $E_{in}(\nu)$ is the electric
field, $\nu$ designates the frequency, $\epsilon_0$ is the vacuum
permittivity and $c$ is the velocity of light. The input optical
pulse is pre-selected to be left-handed circularly polarized, with
polarization vector ${\bf e}_{in}=({\bf x}-i{\bf y})/\sqrt{2}$.  A
polarizing beam splitter (PBS) divides the input pulse into two
orthogonally linearly polarized components with horizontal ({\bf
x}) and vertical ({\bf y}) polarizations, which propagate along
the two arms of a Michelson interferometer. By changing the length
of each arm, $d_1$ and $d_2$, we introduce different time delays
$T_1=2 d_1/c$ and $T_2=2d_2/c$ for each polarization component.
The two delayed pulses recombine at the same PBS. Finally, in the
post-selection stage, the outgoing pulse is projected into a state
of polarization given by the polarization vector ${\bf
e}_{out}=[{\bf x}+\exp(i\Gamma){\bf y}]/\sqrt{2}$, where $\Gamma$
determines the final state of polarization of the output pulse.
For $\Gamma=-\pi/2$, the input and output polarization states
coincide, while for $\Gamma=\pi/2$, they are orthogonal. The
polarization of the output beam  is post-selected with a Liquid
Crystal Variable Retarder (LCVR) (Thorlabs - LCC1113-C) followed
by a polarizer. The relation between post-selection angle and the
LCVR voltage is non-linear and highly temperature dependent. For
this reason, an additional temperature controller is used. After
the polarization post-selection, the electric field of the output
signal writes
\begin{equation}
E_{out}(\omega) =\frac{E_{in}(\omega)}{2} \left[\exp \left(i
\omega T_1 \right) -i\exp \left(i \omega T_2-i \Gamma \right)
\right], \label{eq:output_electric_field}
\end{equation}
where $\omega=2\pi\nu$. Eq. (\ref{eq:output_electric_field}) shows
that the post-selection polarization state ($\Gamma$) determines
for which frequencies the interference between signals coming from
the horizontally and vertically polarized pulses, delayed by
$T=T_1-T_2$, is constructive or destructive.

\begin{figure*}
\includegraphics[width=0.9\textwidth]{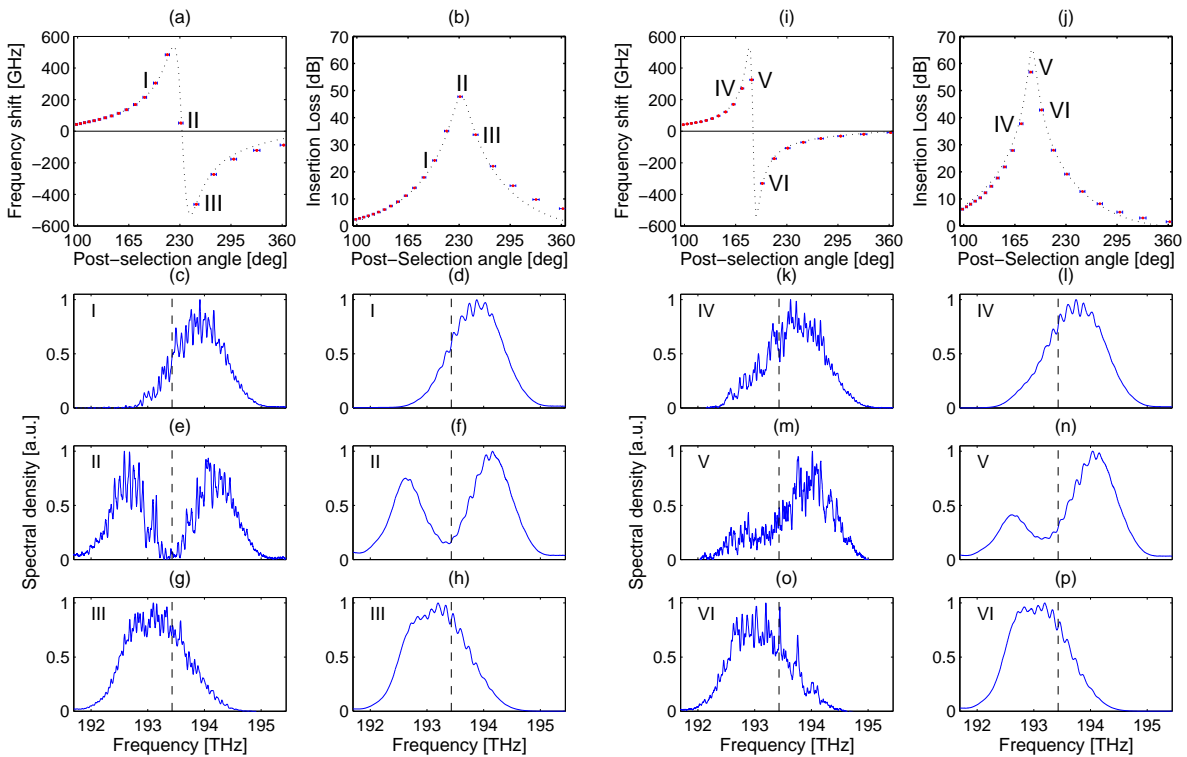}
\caption{Measurement of the central frequency shift induced by
weak value amplification. Measured frequency shift (a) and
insertion loss (b) as a function of the post-selection angle
$\Gamma$. Dots (with error bars) are experimental results, and the
dotted lines are best theoretical fits using the measured input
spectrum in Eq. \ref{eq:SpectrumOutTime}. The best fits are
obtained for $T =53\,\mathrm{fs}$ in (a) and (b), and $T
=22\,\mathrm{fs}$ in (i) and (j). For $T =53\,\mathrm{fs}$: (c),
(e) and (g) (measured) and (d), (f) and (h) (theory) shows the
spectral density for some selected cases, as indicated by the
corresponding labels in (a) and (b). For $T =22\,\mathrm{fs}$:
(k), (m) and (o) (measured) and (l), (n) and (p) (theory) shows
the spectral density for some selected cases, as indicated by the
corresponding labels in (i) and (j). To help the eye, the central
frequency of the input pulse ($\nu_0 = 193.44\,\mathrm{THz}$) is
represented by a dashed line in all plots. The experiment is
performed at a temperature of $34.1^{o}C$. Error bars in all plots
assume that temperature variations during the experiment are in
the range of $\pm 1^{\circ}C$, which translates in random changes
of the angle of post-selection $\Gamma$.}\label{figure2}
\end{figure*}

We measure the output spectral density which is given by
\begin{equation}
S_{out}(\nu) =\frac{S_{in}(\nu)}{2} \left[1+\cos\left(2\pi\nu
T-\Gamma-\pi/2\right)\right]\,. \label{eq:SpectrumOutTime}
\end{equation}
where $S_{in}(\nu)$ is the laser spectrum. In order to
characterize the output spectrum, we measure as a function of the
post-selection angle $\Gamma$, the central frequency shift $\Delta
f=\int d\nu\,\nu \left[ S_{out}(\nu)-S_{in}(\nu) \right]$ and the
insertion loss $L=-10 \log\, F_{out}/F_{in}$, with $F_{in,out}$
being the input (output) energy
$F_{in,out}=\int_{-\infty}^{\infty} S_{in,out}(\nu) d\nu$ of the
pulse. The Optical Spectrum Analyzer (Yokogawa - AQ6370) has a
resolution of $0.02\,\mathrm{nm}$. Each spectrum is obtained after
averaging five data sets in the interval
$[191.5\,\mathrm{THz},195.5\,\mathrm{THz}]$.

Fig. \ref{figure2} shows measurements of the spectral changes in
the regime $T \ll \tau$, when one makes use of the idea of weak
value amplification. It shows the shift of the central frequency
of the spectrum for two different temporal delays: $T=
53\,\mathrm{fs}$ and $T=22$ fs. Fig. \ref{figure2}(a) shows the
measured frequency shift and Fig. \ref{figure2}(b) plots the
measured insertion loss for $T= 53\,\mathrm{fs}$ (similarly Figs.
\ref{figure2}(i) and \ref{figure2}(j) for $T= 22\,\mathrm{fs}$).
The dotted lines are best theoretical fits using the measured
input spectrum in Eq. \ref{eq:SpectrumOutTime}. All other plots in
Fig. \ref{figure2} show measured spectral densities of the output
signal for some selected cases,  and the corresponding theoretical
predictions when the measured input spectral density is used in
Eq. (\ref{eq:SpectrumOutTime}).

Inspection of Fig. \ref{figure2} allows to highlight two working
regimes, corresponding to the presence of high or low losses. For
$\Gamma=-3\pi/2+2\pi \nu_0 T$, there is no central frequency shift
and losses are maximum. The output spectral density features a
double-peak spectral density. For small angle deviations around
this value, central frequency shifts of the spectral density up to
hundreds of gigahertz are clearly observable. However, insertion
losses are also the highest in this regime, measuring values over
$60$ dB. This regime corresponds to the case usually studied in
weak value amplification where the input and output polarization
states are nearly orthogonal \cite{brunner2010}. The applicability
of the weak value amplification in this high-amplification regime
is limited to cases where the energy of the input signal can be
increased, since the intensity of the detected signal is severely
decreased \cite{hosten2008}.

Nevertheless, we demonstrate here that even in the regime where
the input and output polarization states have a significant
overlap---hence featuring smaller insertion losses---weak value
amplification remains useful. Even though the frequency shifts
measured in this regime are generally smaller---reaching only few
tens of GHz instead of hundreds of GHz--- losses do not exceed a
few dB. For $\Gamma=-\pi/2+2\pi \nu_0 T$, there is no shift of the
central frequency again. The pre- and post-selected polarizations
are almost equal, hence introducing almost no losses. The spectral
density of the output pulse is almost equal to the input spectral
density. For small angle deviations around this value, the
temporal delay produce small shifts of the central frequency,
which vary almost linearly with respect to the post-selection
angle. Importantly, these frequency shifts are accompanied by
small insertion losses.

In general, there is a trade-off between the frequency shift
observable for a specific value of the time delay and the amount
of losses that can be tolerated to keep a good signal-to-noise
ratio. The existence of the low-loss working regime, somehow not
so extensively considered as the high-loss regime, can thus
enhance the applicability of the weak value amplification idea, as
demonstrated here.

\begin{figure}
\begin{center}
       \includegraphics[width=0.45\textwidth]{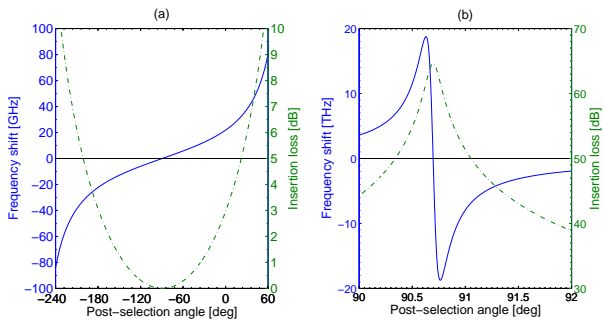}
\end{center}
\caption{Feasibility of the measurement of attosecond temporal
delays with femtosecond pulses. Polarization dependent frequency
shift induced by a $T=10$ time delay of pulses of duration
$\tau=10$ fs. (a) Low-loss and (b) High-loss regime. The solid
line (blue) corresponds to the frequency shift and the dashed
(green) line to the insertion losses. Notice the difference scales
in the $x$ and $y$ axis in (a) and (b).}\label{figure3}
\end{figure}

The results presented here naturally raise the question of what
are the ultimate limits of the scheme, in terms of central
frequency shifts and losses. Brunner et al. \cite{brunner2010} and
Strubi et al. \cite{strubi2013} have estimated theoretically that
weak value amplification of temporally delayed optical pulses
could allow the measurement of attosecond temporal delays. Indeed,
Xu. et al. \cite{xu_guo2013} have demonstrated the measurement of
phase differences as small as $\Delta \varphi \sim 10^{-3}$, which
corresponds to an optical path delay difference of $d=
\lambda/(2\pi)\,\, \Delta \varphi \sim 130$ pm, by using a large
bandwidth LED source. In principle, one can always make use of
white light sources with bandwidths in excess of $100$ nm, as the
ones use in Optical Coherence Tomography for submicron resolution
\cite{pozavay2002}, to enhance the frequency shift detected.

Let us consider as example an input optical pulse with a Gaussian
spectrum, i.e., $S_{in}(\nu) \propto \exp[-\pi^2\tau^2
(\nu-\nu_0)^2 /\ln2]$, where $\tau$ is the pulse temporal width
(FWHM). The central frequency shift $\Delta f$ of the output pulse
can be easily calculated and yields
\begin{equation}
\Delta f = -\frac{\ln2}{\pi}\left(\frac{T}{\tau^2}\right)
\frac{\gamma\sin\big(2\pi\nu_0
T-\Gamma-\pi/2\big)}{1+\gamma\cos\big(2\pi\nu_0
T-\Gamma-\pi/2\big)}\,, \label{FrequencyShiftTime}
\end{equation}
where $\gamma = \exp[-\ln2\,\, T^2/\tau^2]$. The frequency shift
given by Eq. (\ref{FrequencyShiftTime}) is accompanied by
insertion losses which write
\begin{equation}
L =- 10\log\left[\frac{1}{2}\left(1+\gamma\cos\left(2\pi\nu_0
T-\Gamma-\pi/2\right)\right)\right]\,. \label{InsertionLossTime}
\end{equation}
Fig. \ref{figure3} shows the frequency shift expected, as a
function of the post-selection state of polarization, when a $10$
as temporal delay is introduced between two optical pulse with
duration $\tau=10$ fs.

Fig. \ref{figure3}(a) depicts the low-loss regime, where smaller
frequency shifts can be observed in exchange for much lower
losses. In the case shown, frequency shifts up to $100$ GHz,
corresponding to $0.8$ nm, are generated with losses below $12$
dB. Most spectrometers, as the one used in our experiments, can
reach resolutions of up to $0.02$ nm, rendering measurable these
frequency shifts. In the high-loss regime, shown in Fig.
\ref{figure3}(a), one can observe greater frequency shifts, as
high as $\sim 20$ THz ($\sim 160$ nm). Unfortunately, its
measurement is also accompanied by higher losses, over $60$ dB.

In conclusion, we have demonstrated a spectral interference effect
between two optical pulses with a temporal delay much smaller than
the pulse duration, inspired from the concepts of weak
measurements and weak value amplification. In particular, we have
demonstrated a shift of the central frequency of two slightly
delayed femtosecond pulses which can be used to reveal the value
of the temporal delay itself. Importantly, the central frequency
shifts can be observed even in a regime, not so-often considered,
where insertion losses are small, which broadens the applicability
of the method demonstrated.

Our scheme is implemented by using only linear optics elements and
requires spectral measurements, hence making its implementation
practical.The ultimate sensitivity of our scheme can provide
observable frequency shifts for temporal delays of the order of
attoseconds using femtosecond laser sources. Our scheme thus
appears as a promising method for measuring small and rapidly
varying temporal delays.

\vspace{0.2mm} \noindent \small{{\bf Acknowledgements}: We
acknowledge support from the EU project PHORBITECH (FET OPEN grant
number 255914), the Spanish government projects FIS2010-14831,
TEC2010-14832, the Ramon y Cajal and Severo Ochoa programs, and
Fundaci\'o Privada Cellex, Barcelona. LJSS acknowledges support
from the ``Fundaci\'{o}n Mazda para el Arte y la Ciencia'',
Bogot\'{a}, Colombia, and NB from the Swiss National Science
Foundation (grant PP00P2\_138917).}

\end{document}